\documentclass{aastex}

\begin{document}

\title{Generation of nanoflares in the Crab pulsar}
%\slugcomment{Not to appear in Nonlearned J., 45.}
%% Running heads
%\shorttitle{Nanoflares in Crab pulsar}
%\shortauthors{Machabeli et al.}

\author{G. Machabeli}
\affil{Institute of Theoretical Physics,
Ilia State University, 3 ave. Cholokashvili, Tbilisi 0162,
Georgia}
%\email{\emaila}

\author{I. Malov}
\affil{Pushchino Radioastronomical Observatory, 142290 Moscow Region, Pushchino, Russia}

\and

\author{G. Gogoberidze}
\affil{Institute of Theoretical Physics,
Ilia State University, 3 ave. Cholokashvili, Tbilisi 0162,
Georgia}

%\altaffiltext{1}{First Alternate Affilation.}
%\altaffiltext{2}{Second Alternate Affilation.}
%\altaffiltext{3}{Third Alternate Affilation.}

\begin{abstract}
We study dynamics of drift waves in the pair plasma of pulsar magnetosphere. It is shown that nonlinear of the drift waves with plasma particles leads to the formation of small scale structures. We show that cyclotron instability developed within these nonlinear structures can be responsible for the formation of  nanoshots discovered in the radio emission of the Crab pulsar.
\end{abstract}


%\section*{}
%\label{sec:intro}

\section{Introduction}

The Crab nebula is a distinctive object in our galaxy. One of its the puzzling feature is related to the emission of the pulsar PSR 0531+21 in the crab nebula. Namely, it is long known \citep{MT} that radio, optical, $X$ and $\gamma$-ray emission of this pulsar comes from the same region in the magnetosphere which should be located close to the light cylinder (hypothetical surface where rotational speed of magnetic field lines becomes equal to the speed of light). Explanation of this fact is one of the challenges of any model of the pulsar radio emission. There exist several models of pulsar radio emission \citep{RS75,M78,BEA88,KMM91,W97} but up to the date non of them can entirely explain all the observations.

One possible mechanism of the pulsar radio emission is similar to the mechanism responsible to the solar radio bursts. Two stream instability between primary electron beam and pair plasma of the pulsar magnetosphere causes generation of strong Langmuir turbulence \citep{W97}. This turbulence cannot escape the plasma directly. The modulational instability in strong Langmuir turbulence generates solitons and nonlinear wave-wave coupling within the solitons generates bursts of electromagnetic radiation which can escape the plasma.

The electron maser emission mechanism also implies that the beam of plasma generates Langmuir turbulence. When the turbulence becomes strong, both the driving beam and turbulent fluctuations will become spatially inhomogeneous \citep{BEA88}. As a result, the collective motion of charge bunches interacting with electrostatic fluctuations in the turbulence results in coherent, forward-beamed radiation.

Another possible mechanism of the radio emission is related to the development of cyclotron instability \citep{KMM91,LBM99}. The cyclotron instability develops when the condition of anomalous Doppler effect resonance is fulfilled. This instability generates electromagnetic waves which directly escape the plasma without need of the mode conversion.

Another possible mechanisms include radiation caused by particle acceleration in the oscillations electric field parallel to the magnetic field of a pulsar \citep{M78}, curvature emission by coherent bunches \citep{RS75} and curvature maser emission \citep{LM92}.

It is not an easy task to compare predictions of the different models to observations. The main reason is that all the models deal with microscopic plasma processes in the magnetosphere whereas most of the existing data is linked to the large scale processes. Recently, very high time resolution observations allowed \citet{EH16} to compare data of the nanoshots observed in the main pulse and low-frequency inter-pulse of the pulsar PSR 0531+21 with predictions of the different models of the pulsar radio emission. The authors concluded that non of the existing models can explain three main characteristics of their observations: spectrum, timescale of the phenomenon and observed significant level of circular polarization. Regarding the model based on the development of the cyclotron instability \citet{EH16} concluded that this model naturally explains observed significant level of circular polarization \citep{KMM91,GM5}, but on the other hand, because the generation mechanism is linear, in the basic version of the model there is no explanation for the existence of the phenomenon with characteristic time scale of nano-seconds.

In the presented paper we study development of the cyclotron instability in the pair plasma of the pulsar magnetosphere and will show that interplay of the instability development with the drift waves can lead to the formation of nanoshots. The paper is organized as follows: In Section 2 we study generation the electromagnetic waves in the pair plasma of the pulsar magnetosphere by means of cyclotron instability. The origin of the drift waves in the magnetosphere is studied in section 3. Mechanism of the nanoshot formation is presented in Section 4. Conclusions are given in Section 5.

\section{Development of the cyclotron instability}

As in the standard model of pulsar magnetosphere we assume that the relativistic electron-positron plasma is moving along the magnetic field lines. In the case $n^{-1/3}<\lambda$, where $n$ is plasma density and $\lambda$ is the wavelength of a wave, it is necessary to take into consideration the effect of wave interference (i.e., collective behavior of the particles). \citet{B75} showed that Cherenkov radiation in the pair plasma is suppressed. But later it was shown \citep{KMM91,LBM99} that generation of the electromagnetic waves is possible at modified Cherenkov resonance (so-called Cherenkov-drift resonance) if drift motion of the particles caused by the curvature of the magnetic field is taken in to account. The condition of the Cherenkov-drift resonance is
\begin{equation}
(\omega - k_\parallel v_\parallel- k_\perp u_\perp)=0.  \label{1}
\end{equation}
Here $v_\parallel$ is the parallel component of the particle velocity,  $u_\perp =\gamma v_\parallel^2/\omega_B R_c$ is the electron drift speed caused by curvature of the field line, $R_c$ is the curvature radius of the field line, $\omega_B=eB_0/mc$ is the cyclotron frequency, $e$ is an elementary charge, $m$ is the electron mass, $c$ is the speed of light, $B_0$ is the magnetic field of the pulsar, $\omega$ is the wave frequency, $k_\parallel$ and $k_\perp$ are components of the wave vector parallel and perpendicular to the magnetic field, respectively.

The particles drift in the direction perpendicular to the curved magnetic field lines.  The drift of particles cause brake of the axial symmetry. Mathematically this means that in the dispersion relation of electromagnetic waves in addition to the perpendicular component of the electric field $E_\perp$ there also appear terms proportional to the parallel component $E_\parallel$. The later terms contain resonances given by equation (\ref{1}). \citet{GGM02} considered generation of radio waves in the approximation of infinite magnetic field $B_0 \rightarrow \infty$. In this research drift motion was ignored but inclined propagation (with respect to the magnetic field) was assumed and this also caused the brake of the axial symmetry.

The particle distribution function of the electron-positron plasma in the pulsar magnetosphere describes the plasma moving from the surface of the star to the light cylinder and consists of three components \citep{GJ69,S71,T73}. The first one is the primary beam of electrons with Goldreich-Julian density $n_b=n_{GJ}=7\times10^{-2}B_\star/P$, where $B_\star$ is the magnetic field on the surface of the pulsar and $P$ is the spin period.
Typical Lorentz factor factor of the primary beam is $\gamma_b\sim 10^{6}-10^{7}$; 2. So-called tale particles \citep{T73} with $n_t\sim 10^{13}-10^{15}~{\rm cm^{-3}}$ and typical Lorentz factor $\gamma_t\sim 10^{3}-10^{5}$; 3. The bulk pair plasma. The density of the bulk plasma strongly depends on the character of the pulsar magnetic field. If the magnetic field of the pulsar is dipolar, then the density $n_p\approx \gamma_b n_b/\gamma_p \sim 10^{18}~{\rm cm^{-3}}$, and  $\gamma_p\sim 10^{2}$ is Lorentz factor of the bulk pair plasma. On the other hand, for quadrupolar magnetic field \citep{MU89} we have $\gamma_p\sim 3-10$ and $n_p \sim 10^{20}~{\rm cm^{-3}}$. The entire particle distribution function is one dimensional. This kind of distribution function is unstable with respect to the cyclotron instability \citep{KMM91,LBM99} when the condition for the anomalous Doppler resonance is fulfilled
\begin{equation}
\omega -k_\parallel v_\parallel -k_\perp u_\perp + \frac{s\omega_B}{\gamma_{res}}=0.   \label{2}
\end{equation}
Here $s=\pm1,\pm2,\pm3...$ is the number of the cyclotron frequency harmonics and $\gamma_{res}$ is the Lorentz factor of the resonant particles.

Effective generation of the radio waves takes place in the vicinity  of the light cylinder. The generated waves propagate along the tangent of the field line. Therefore an observer should detect the emission when the line of sight is parallel to the field lines.

For the analysis of the resonant condition (\ref{2}) we need to know main characteristics of the eigenmodes of the electron-positron plasma. These exist three electromagnetic eigenmodes of the pair plasma which propagate along the field lines and can escape from the magnetosphere. One of the eigenmodes, so-called $O$-mode is purely transverse. Its electric field $E_{O}$ is perpendicular to the plane formed by the wave vector and the background magnetic field. Two other eigenmodes (so-called $A$ and $X$-modes) are mixed longitudinal-transverse waves. Their electric field is placed in the plane formed by the wave vector and the background magnetic field. Simple dispersion relations for the different modes can be found for the waves propagating almost parallel to the field lines. In this case in the laboratory frame the dispersion relations of $O$, $A$ and $X$-modes are, respectively
\begin{equation}
\omega_X=kc(1-\delta),~~~{\rm where}~~~\delta=\frac{\omega_p^2}{4\gamma_p^3 \omega_B^2} \label{3}
\end{equation}
\begin{equation}
\omega_{A}=k_\parallel c \left( 1-\delta - \frac{k_\perp^2 c^2}{4\gamma_p \omega_p^2} \right), \label{4}
\end{equation}
\begin{equation}
\omega_{O}^2= \omega_p^2 \gamma_p^{-3} + k_\parallel^2 c^2. \label{5}
\end{equation}
In equation (\ref{3}) $\omega_p^2=8\pi e^2n_p/m$ is the plasma frequency.

The standard model of pulsar magnetosphere implies that electrostatic field is generated at the polar caps of the pulsar. This electric field pulls out electrons from the surface of the star. In the strong magnetic field of the pulsar (at the surface of the pulsar PSR 0531+21 the magnetic field $B_\star \approx 7.6\cdot10^{12}~{\rm G}$), transverse component of momentum of primary electrons is emitted by the synchrotron radiation for the characteristic timescale $10^{-15}~{\rm s}$ and the electrons reach Landau's zeroth level. Later the primary beam generate electron-positron plasma through avalanche process \citep{S71}. Thus, the particle distribution function in the magnetosphere is one dimensional. This kind of distribution function is unstable with respect to the cyclotron instability and the generation of radio waves can take place when the resonant condition (\ref{2}) is fulfilled.
During the quasi-linear stage of the instability the radio waves react back on the particle distribution function and this causes diffusion of particles both along and perpendicular to the field lines. This circumstance leads to the saturation of the instability. On the other hand, particles gain transverse components of momentum and due to the synchrotron radiation generate electromagnetic waves in the $X$ and $\gamma$ bands. This scenario explains how radio and high frequency waves can be generated at the same location, thus explaining coincidence of the peaks of pulses in different frequency bands.

The frequency of the generated radio waves can be found from equation (\ref{2}) using equation (\ref{3}) and the expansion
\begin{equation}
v_\parallel \approx c \left( 1- \frac{1}{2\gamma_{res}^2} \right),~~~{\rm and}~~~k\approx k_\parallel \left( 1+ \frac{k_\perp^2}{2k_\parallel^2} \right). \label{6}
\end{equation}
In this case
\begin{equation}
\omega\approx k_\parallel c = \frac{s\omega_B}{\gamma_{res} \delta} = 4\gamma_p^3 \frac{s\omega_B^3}{\gamma_{res} \omega_p^2}   . \label{7}
\end{equation}
Characteristic frequency of the high frequency synchrotron radiation is
\begin{equation}
\omega\approx \omega_B \gamma_{res}^2. \label{8}
\end{equation}
The last two equations give relation between generated radio and high frequency waves.

According to equation (\ref{7}) for $\gamma_p \approx 3$ the radio waves with frequencies about several GHz can be excited by resonant particles with the Lorentz factors $\gamma_{res} \sim 10^{7}$. Using these parameters for the quadrupole magnetic field \citep{MU89} we have $\omega_p^2 \approx 6.4\cdot 10^{28} (R_\star/r)^4~({\rm rad/s})^2$ and $\omega_B \approx 1.4\cdot 10^{20} (R_\star/r)^4~{\rm rad/s}$. Using equation
(\ref{7}) and assuming that the generation of radio waves take place at $r\sim 10^{8}~{\rm cm}$ we obtain for the frequency of the radio waves
\begin{equation}
\omega \approx 4\cdot 10^{9}~{\rm rad/s}. \label{13}
\end{equation}
For $s=1$ the radio emission corresponds to the frequencies of the order of 1 GHz. Waves with frequencies $\sim 0.1~{\rm GHz}$ are generated by resonant particles with $\gamma_{res}\sim 10^{8}$. There are very few particles with such a high Lorentz factor and therefore one can expect that at these frequencies intensity of radiation should be significantly reduced.

Because the particles are moving along the field lines to the observer, angular distribution of the radiation reaching the observer is strongly anisotropic \citep{SEa76,LL} and is mainly concentrated within the angle
\begin{equation}
 \alpha \approx \frac{1}{\gamma_{res}} \label{12}
\end{equation}
around the field line.

\section{Generation of the drift waves}

In the previous section we considered the waves propagating almost parallel to the magnetic field. Here we consider possibility of $A$-wave generation with wave vectors almost perpendicular to the magnetic field \citep{KMM91,MKMS,GMML,MM}. Plasma particles moving along the  field lines are subject to the drift motion due to the curvature of the field lines. The drift speed is
\begin{equation}
u_\perp =\frac{\gamma v_\parallel^2}{\omega_B R_c}. \label{14}
\end{equation}
The particles are drifting in the direction perpendicular to the plane containing the curved field lines. It is convenient to study interaction of the drift wave with plasma particles in the local cylindrical frame of reference $(x,r,\phi)$. $x$-axis is directed perpendicular to the plane containing the curved magnetic field line. ${\bf r}$-direction is perpendicular both to the $x$-axis and the field line. The center of the system is placed at the center of curvature of the field line and consequently the angle $\phi$ determines location of the particle along the magnetic field line.

\citet{KMM91b} studied generation of almost transverse ($k_x/k_\phi\gg 1$) drift waves at the modified Cherenkov resonance (\ref{1}).
For the frequency of the generated wave equation  (\ref{1}) yields
\begin{equation}
\omega \approx k_x v_x+ k_\phi u_\phi,  \label{15}
\end{equation}
whereas for the maximum growth rate wa have \citep{KMM91b,GMML}
\begin{equation}
\Gamma \approx \left( \frac{n_b \gamma_p^3}{n_p \gamma_b} \right)^{1/2} k_x u_x.  \label{16}
\end{equation}

The growth rate (\ref{16}) has maximum value for the waves with wave vectors $k_x^2 \ll 3\omega_p^2/2\gamma_p^3 c^2$. For parameters of the crab pulsar this condition gives $k_x \ll 0.5 {\rm cm^{-1}}$  and determines wave vector of the excited waves. Using the condition $n_p\gamma_p \lesssim n_b \gamma_b$ for typical value of the grows rate for the crab pulsar equation (\ref{16}) gives $\Gamma \sim 10 {\rm sec^{-1}}$.

 The generated drift waves are propagating almost perpendicular to the magnetic field. They circulate around the magnetic field lines, slowly approaching the surface of the light cylinder following spiral path. Therefore these waves stay quite a long time in the magnetosphere and can effectively participate in various dynamical processes. The time of propagation of the drift waves to the light cylinder is $k_x/k_\phi$ times greater compared to the escaping time of particles from the magnetosphere. Let us estimate the amplitude of generated drift waves. The energy source of the instability is kinetic energy of the particles drifting in the inhomogeneous magnetic field. Therefore the maximum amplitude $B_r$ of the drift waves can be estimated comparing energy of the drifting particles and energy of the waves. This condition yields
 \begin{equation}
mu_x^2n_b\gamma_b \tau_e \frac{k_x}{k_\phi} \sim \frac{B_r^2}{8\pi}.  \label{16a}
\end{equation}
Here $\tau_e$ is escaping time of the particle from the magnetosphere. This equation shows that for $k_x/k_\phi \sim 10^9$ energy density of the generated magnetic field at light cylinder can become comparable to the energy of pulsars magnetic field.

Drift waves can be responsible for the observed so-called zebra structures in the pulsar radio emission \citep{EH16}. Indeed, existence of the drift waves makes magnetic field structure locally inhomogeneous and therefore in different places condition for Anomalous Doppler effect (\ref{2}) can be fulfilled for different harmonics $s=0, \pm1, \pm 2$ \citep{Zea}. The grows rate of electromagnetic waves is different for different harmonics and consequently the electromagnetic waves would have different amplitudes.

\section{Nonlinear processes and formation of nanoshots}

Let us consider the drift wave propagating almost perpendicular to the magnetic field, satisfying condition $k_xu_x \gg k_\phi v_\phi$. For such waves we have $\omega \approx k_xu_x$. From Maxwell equation we have $B_r =E_\phi (k_xc/\omega) \sim E_\phi (c/u_x)$, where ${\bf B}$ and ${\bf E}$ are perturbations of the magnetic and electric fields, respectively. From this equation it follows that $B_r \gg E_\phi$. Curvature of the field line $\rho_c=1/R_c$ in the Cartesian frame of reference is defines by the equation $ (1+ dy^2/dx^2)^{-3/2}(d^2y/dx^2) dy/dx=B_y/B_x$, and using equation $(\nabla \cdot {\bf B})=0$, in the cylindrical frame we have
\begin{equation}
\rho_c=\frac{1}{rB} \left[B_\phi - \frac{B_\phi^2}{B^2} \frac{\partial B_r}{\partial \phi} \right],  \label{17}
\end{equation}
where $B=(B_\phi^2+B_r^2)^{1/2}\approx B_\phi(1+B_r^2/2B_\phi^2)$. In the azimuthal direction we have $B_r \sim \exp(ik_\phi r) =\exp(iJ\phi)$, where $J \equiv k_\phi r/\phi$. For $J\gg 1$ equation (\ref{17}) yields
\begin{equation}
\rho_c=\frac{1}{R_B} \left(1 - J \frac{B_r}{B_\phi} \right).  \label{18}
\end{equation}
For high harmonics ($J \gg 1$) the drift wave can significantly change the curvature of the magnetic field.
Let us consider how this change will influence resonance conditions (\ref{1}) and (\ref{2}). Using expansion $v_\phi \approx c(1-1/2\gamma^2-u_\perp^2/2c^2)$ and dispersion relation for $O$-mode, $\omega_{O} =kc(1-\delta)$, equations (\ref{2}) yields
\begin{equation}
\frac{1}{2} \left[ \frac{k_x}{k_\phi} - \frac{u_\perp}{c}\left( 1- J\frac{B_r}{B_\phi}\right) \right]+\frac{k_r^2}{4k_\phi^2}-\delta=-s\frac{\omega_B}{\gamma_r k_\phi c}.  \label{19}
\end{equation}
This equation describes both the anomalous Doppler resonances ($s=+1,+2 ...$) and the Cherenkov resonance ($s=0$). Due to the presence of the drift waves neither the magnetic field nor its curvature are constant. Therefore, the resonant condition (\ref{19}) can be fulfilled only for special areas, thus forming 'emitting spots' in the magnetosphere. In the areas where the drift wave causes increase of the magnetic field curvature there is increase of the magnetic field as well. Because the relativistic particles flow along the field lines, increase of the field causes increase of particle concentrations and as a result enhancement of the radiation intensity. Characteristics of the 'emitting spot' strongly depends on the phase of the drift wave as well as its longitude and latitude \citep{MKMS}.

In the weak turbulence approximation nonlinear dynamics of the drift waves in the pulsar magnetosphere can include 3 and 4-wave resonant interaction and nonlinear interaction of waves with plasma particles \citep{GMML}. A three-wave interaction involves quadratically nonlinear current which is an odd function of the sign of the charge, so that electrons and positrons contribute with opposite sign, making this effect relatively weak in a pulsar pair plasma \citep{LM94}. Four-wave interaction, which involve the cubic nonlinearity, is intrinsically small, and nonlinear interaction of waves with plasma particles appears the strongest nonlinear process involving the drift waves \citep{GMML}. This interaction includes both scattering of waves by plasma particles and merging of waves. In the former case the wavelength increases and this process is limited by the dimensions of the magnetosphere, whereas in the case of the wave merging the wavelength decreases.

Study of the nonlinear interaction of waves with plasma particles in the electron-ion plasma involves consideration of a test particle \citep{GS} which causes polarization of nearby media (forming so-called shielding cloud). In contrast, in the pair plasma polarization contributions of electrons and positrons cancel out \citep{M83} and therefore effects related to the polarization are absent. Conservation of energy and momentum for the the nonlinear interaction of waves with plasma particles leads to the following conditions
\begin{equation}
\omega \pm \omega^\prime= (k_\phi \pm k_\phi^\prime)v_{\parallel}.  \label{21}
\end{equation}
Here primed quantities belong to the wave after the nonlinear interaction. The minus sign in equation (\ref{21}) corresponds to the nonlinear scattering of the waves and the plus sign describes the merging of the waves. The nonlinear scattering of the drift waves has been intensively studied before \citep{KMM91,MKMS,GMML}. This process transfers the wave energy to the particles and therefore energy cascades to the lower frequencies, i.e., larger scales. On the contrary, the merging of the waves increase the wave frequency and transfers the energy to smaller scales. Time scales of these processes are of the same order of magnitude \citep{GS} and as shown before \citep{MKMS,GMML} the energy can be effectively transferred to different scales before the drift waves leave the magnetosphere. Therefore, merging of the waves leads to the formation of small scale drift wave structures. Intensity of the magnetic field as well as plasma density in these structures is enhanced within these structures and therefore we have small scale intensively emitting spots.

If the size of the emitting spot is negligible (compared to the size of the magnetosphere) then duration of its emission received by an observer is (see equation (\ref{12}))
\begin{equation}
\tau \approx \frac{1}{\Omega \gamma_{res}}.  \label{22}
\end{equation}
Noting that for the crab pulsar $\Omega \approx 200$ and for the particles of primary beam $\gamma \sim 10^6-10^7$ minimal duration of the signal associated with a small emitting spot is $\tau \sim 10^{-9}$ s. As we see, characteristic timescale corresponds to the observed timescale of the nanoshots.

We suggest that due to the nonlinear evolution of the drift waves at the same time many emitting spots are formed in the pulsar magnetosphere, but only few of them are crossing observer's line of sight.

\section{Conclusions}

In the presented paper we studied influence of the drift waves on the formation of radio emission in the pair plasma of the pulsar magnetosphere. It was shown that nonlinear evolution of the drift waves are dominated by the nonlinear interaction of waves with plasma particles which include both induced scattering and merging of the waves. The latter process transfers the energy to the smaller scales and leads to the formation of small scale drift wave structures where magnetic field and plasma density are enhanced. These structures form  intensively emitting spots. Duration of the signal associated with a spot is of the order of nano-seconds and consequently the drift wave structures can be responsible for the recently discovered nanoshots in the radio emission of the Crab pulsar. Our model implies that at the same time many emitting spots are formed but only few of them cross the observer's line of sight.

Because we suggest that different nanoshots are related to the different emitting spots, our model predicts that there should be no correlation between various characteristics (such as circular polarization, etc.) of different nanoshots.

\acknowledgments
This work has been supported by Shota Rustaveli National Science
Foundation grants FR/51/6-300/14 and FR/516/6-300/14.

\end{document}